\documentstyle[twocolumn,epsf,aps]{revtex}
\begin{document}
\title{On quantum key distribution in decoherence-free 
subspace
}
\author{Xiang-Bin Wang\thanks{email: wang$@$qci.jst.go.jp} 
\\
        Imai Quantum Computation and Information project, ERATO, Japan Sci. and Tech. Corp.\\
Daini Hongo White Bldg. 201, 5-28-3, Hongo, Bunkyo, Tokyo 113-0033, Japan}

\maketitle 
\begin{abstract}
We propose easy implementable protocols for robust quantum 
key distribution with
the collective dephasing channel or collective  rotating channel. 
In these protocols, Bob only takes passive photon detection to
measure the polarization 
qubits in the random bases. The source for the protocol
with collective rotating channel is made by type 2 spontaneous parametric
down conversion with random unitary rotation and phase shifter, no
quantum disentangler is required.
A simple proof for unconditionally security is shown.
\end{abstract}
Consider the following collective dephasing channel to polarization
qubits\cite{ekert}:
\begin{eqnarray} \nonumber
|0\rangle\longrightarrow e^{i\phi_0}|0\rangle;
|1\rangle\longrightarrow e^{i\phi_1}|1\rangle\\
 |0\rangle|1\rangle\longrightarrow e^{i\Delta}|0\rangle|1\rangle;
|1\rangle |0\rangle\longrightarrow e^{i\Delta}|1\rangle |0\rangle
\label{dephase}
\end{eqnarray}
and $\Delta=\phi_0+\phi_1$. Here $|0\rangle,|1\rangle$ represent for horizontal
and vertical polarization states respectively.
Obviously, 2-qubit state $|01\rangle,|10\rangle$ and their arbitrary linear superposition will be robust
 via the above defined noise, since they will only
obtain an unimportant overall phase shift which is not detectable.
Naturally, one may consider to replace the original
single qubit BB84 states \cite{BB} by the 2-qubit states randomly chosen
from the set $\{|01\rangle,|10\rangle,\frac{1}{\sqrt 2}(|01\rangle+|10\rangle),
\frac{1}{\sqrt 2}(|01\rangle-|10\rangle)\}$ to do the quantum key distribution (QKD)\cite{BB,gisin}
if the dominant  channel noise is the above defined 
collective dephasing noise.   There has been a proposal on how to 
experimentally realize the decoherence-free subspace QKD very recently
\cite{walton,boil}.
However, no security proof for any of such type of protocols is given so far.
Also, the existing protocol\cite{boil} with collective rotating error
uses 4-photon state which is technically very difficult in practice.
In this paper, we propose easy implementable robust QKD 
protocols with both collective dephasing channel\cite{walton} and
collective rotating channel\cite{boil} and we also give a simple proof
for the unconditional security in both cases.
We first give a QKD protocol that tolerates the collective phase error
as the following:
\\{\it protocol 1}
{\bf 1:} Alice creates $(4+\delta)n$ single qubit state
randomly chosen from
$\{|0\rangle,|1\rangle,|+\rangle,|-\rangle \}$ 
($|\pm\rangle=\frac{1}{\sqrt 2}(|0\rangle\pm|1\rangle)$) and 
$(4+\delta)n$ ancilla which are all in state $|0\rangle$.
She then encodes each individual qubit with an ancilla 
into a 2-qubit code through
$|00\rangle\longrightarrow |01\rangle; |10\rangle\longrightarrow |10\rangle$
where the second qubit states in the left hand-side of the arrows are for the
ancilla.
After this operation, she now has made  $(4+\delta)n$
2-qubit quantum codes with each of them randomly chosen from the set $\{|01\rangle,|10\rangle,\frac{1}{\sqrt 2}(|01\rangle+|10\rangle), \frac{1}{\sqrt 2}(|01\rangle-|10\rangle)\}$.
{\bf 2:} For each 2-bit code, she puts down the ``preparation basis'' as 
``Z basis''
($\{|0\rangle,|1\rangle\}$ basis) if
it is in state $|01\rangle$ or $|10\rangle$ and the preparation basis
as ``X basis'' ($\{|\pm\rangle\}$) if it is in one of the states $\{\frac{1}{\sqrt 2}(|01\rangle+|10\rangle), \frac{1}{\sqrt 2}(|01\rangle-|10\rangle)\}$. 
For those code states of  $|01\rangle$ or  
$\{\frac{1}{\sqrt 2}(|01\rangle+|10\rangle)$, she denotes a bit value 0 ;
for those code states of $|10\rangle$ or  
$\{\frac{1}{\sqrt 2}(|01\rangle-|10\rangle)$, she denotes a bit value 1.
{\bf 3:} Alice sends the 2-qubit codes to Bob.
{\bf 4:} Bob receives the $(4+\delta)n$ 2-qubit codes. To each code,
He measures 
the first qubit (the one in the right position of the state) in $X$ basis and then measures the other qubit
(say, qubit 2) in either X basis or Z basis 
after taking a unitary 
transformation U to it. Unitary U to qubit 2 is dependent on the 
measurement outcome of qubit 1: If the outcome is $|+\rangle$, U is unity, $I$;
if it is $|-\rangle$, U is $\sigma_z$ which is
defined by $\sigma_z|0\rangle=|0\rangle; \sigma_z|1\rangle=-|1\rangle$. 
Bob denotes his ``measurement basis'' just the same as his measurement basis
to qubit 2. And, if the outcome to qubit 2 is $|0\rangle$ or $|+\rangle$, he puts down a bit value 0; if the outcome to qubit 2 is $|1\rangle$ or $|-\rangle$, he puts down a bit value 1.
{\bf 5:} Alice announces her ``preparation basis'' for each codes.
{\bf 6:} Alice and Bob discards those bits on which Bob has
 measured in a basis
       diffrent from Alice's preparation basis.  
         With high probability, there are at
	least $2n$ bits left (if not, abort the protocol).  Alice 
        decides randomly on a set of $n$ bits to use for the protocol, and
        the rest to be check bits. 
{\bf 7:} Alice and Bob announce the values of their check bits.
	If too few of these values agree, they abort the protocol.
{\bf 8:} Alice and Bob distill the final key by using the classical
CSS code\cite{shorpre}.\\
Before we make the security proof, we first take a look at its fault tolerance
property. Since we have already assumed that the main error of the physical 
channel is the collective dephasing error as defined by eq.(\ref{dephase}),
all the 2-qubit codes will be sent robustly over the physical channel, since
the collective dephasing error will now only offer a trivial overall
phase factor $e^{i\Delta}$. After Bob receives the 2-qubit codes, he decodes them in step 4. 
One may check it easily that after decoding, the original BB84
states are recovered provided that Bob's measurement basis to qubit 2
is same with Alice's
``preparation basis''.
\\
We now give a very simple security proof for the above protocol in the way 
that if the protocol above is insecure, then the known
BB84 protocol with CSS code
proposed by Shor and Preskill\cite{shorpre} is also insecure.
We first compare our protocol here with Shor-Preskill protocol\cite{shorpre}.
The only difference between them is the encoding and decoding in our protocol.
Without these encoding and decoding, our protocol is exactly just 
Shor-Preskill protocol. Suppose our protocol is insecure then eavesdropper, Eve
must have a certain intercept-and-resend scheme S to obtain significant 
information to the final key in our protocol. 
Most generally, S may contain the intercepting
the qubits from Alice,
operation $\hat A$ to all qubits and ancilla, resending the qubits to Bob
and finally a certain operation $\hat O$ to the rest qubits with Eve. 
Operation $\hat O$ can include certain type of measurement which optimizes
Eve's information to the final key.
We now show that if scheme S can help Eve to obtain significant information to
the final key of our protocol, then we can always construct a scheme S' for
Eve to obtain the same amount of information to the final key of
Shor-Preskill protocol.

Since the encoding procedure in our protocol requires no 
information to the original single qubit state  and the decoding procedure
requires no information to the 2-qubit code state, the encoding and decoding
can actually be done by {\it anybody}. Lets consider the Shor-Preskill 
protocol. Eve now uses attack S'. In scheme S', Eve  replaces operation
$\hat A $ by $$
Encoding\longrightarrow\hat A\longrightarrow Decoding
$$
and everything else     identical to that in scheme S. Note that in scheme S',
Eve resends the decoded single qubits to Bob and discards all ``qubit 1'' 
in each code after decoding. If Eve uses this S' to Shor-Preskill's protocol,
everything will be identical with that in our protocol with Eve's attack
S. In other words, to Eve, the game of attacking Shor-Preskill protocol with
scheme S' is exactly identical to the game of attacking our protocol
with S since she may just regard her encoding and decoding as operations
done by someone else.
More specifically, given Shor-Preskill protocol with Eve's
attack S' and our protocol with Eve's attack S,
everything of the two protocols are identical therefore Eve's information
to the final key of the two protocols must be exactly the same. 
This is to say, if our protocol is insecure, Shor-Preskill protocol must be 
also insecure. However, the security of Shor-Preskill protocol has been
proven\cite{shorpre} already, therefore our protocol must be secure.

Actually, we can give a more general theorem about the security of
any protocol with quantum code. 
Consider two protocols, protocol P0 and protocol P. In protocol P0,
 Alice directly sends Bob
each individual qubits. In protocol P, Alice first encodes each individual qubits by a certain
quantum code and then sends each quantum codes to Bob. Bob will  decode the transmitted codes. 
After decoding, Bob obtains one single qubit state from each code and 
discarded all other qubits in the same code. Alice and Bob continue
the protocol. Suppose except for the extra steps of encoding 
and decoding, everything
else in protocol P0 and protocol P are identical and the encoding and decoding
do not require any information of the quantum state of the original qubit or the codes, 
then we have the following theorem:
{\bf If protocol P0 is secure then protocol P is also secure.}
The proof can obviously be done by the similar arguement: Eve can also do
the encoding and decoding, if P is insecure under attacking scheme S, 
we can always
construct an attack S' for Eve to attack P0 effectively.

Note that there are two important points here. First, the encoding must be independent
of the original quantum state itself therefore Eve can also do so. Also this shows
that the encoding does not offer extra information to the qubit state.
The simple repetition code does not work because the encoding needs the information to the 
original quantum state and Eve is not able to do so. Also the repetition code will offer more information
to the bit value. Secondly, Alice and Bob test the error rate $after$ the decoding, that is to say,
they directly test the error rate to the decoded qubits instead of testing the error rate
of each qubits before decoding and then mathematically deduce the error rate after decoding by using
the collective dephasing property of the physical channel. Note that Eve may change all properties
of the physical channel when the QKD task is being done. But in our protocol the error test is done
after the decoding, this will defeat any type of Eve's attack. Since Eve does not want to be detected,
 she has to consider the error test done by Alice and Bob latter on. Therefore she must limit herself
to those types of attack which will not affect the result of error test 
to be done by 
Alice and Bob. (Here Eve does not have to respect the error rate of the physical channel since that rate
will not be tested in the protocol, but Eve has to respect the error rate after decoding since this is the issue that will be tested.) That is to say, {\it
if Eve is not detected by the error test, the protocol will work as
effectively as the case that there is no Eve}, since the 
key rate of the final key is {\it uniquely} determined by 
the error rate in test. 

After the above security proof, we can now further simplify our protocol.
The encoding procedure in step 1 is just the local operation by Alice.
She does not have to really first prepare the single qubit state and then encodes
it by controlled-not (CNOT) operation. Instead, she may directly prepare the 2-qubit codes
with each of them randomly chosen from $\{|01\rangle,|10\rangle,
\frac{1}{\sqrt 2}(|01\rangle + |10\rangle),\frac{1}{\sqrt 2}(|01\rangle - |10\rangle) \}$ . These random codes can be produced with type II 
spontaneous parametric
down conversion (SPDC)\cite{para} with a certain filter\cite{walton}.
In step 4 of our protocol, Bob  measures qubit 2 in either X basis or Z 
basis. This requires Bob to change measurement basis rapidly. However, this 
task can be done by passive detection with a 50:50 beam splitter, see figure 
1. In step 4, Bob  takes unitary transformation 
U to qubit
2 according to the measurement result of qubit 1. However, since we have already limitted 
ourselves to the 4 BB84 states, instead of taking U before the measurement to qubit 2, he may directly measure qubit 2 and then determine the
corresponding bit value $b$ according to the measurement result of both qubit 1
and qubit 2 through the following rule:
\begin{eqnarray}
\nonumber \{|+\rangle|0\rangle,|-\rangle|0\rangle,|+\rangle|+\rangle, 
|-\rangle|-\rangle\}\longrightarrow b=0;
\\ \{|+\rangle|1\rangle,|-\rangle|1\rangle,|+\rangle|-\rangle, |-\rangle|+\rangle\}\longrightarrow b=1.
\label{corres}
\end{eqnarray}
Note that the state preparation and the bit value detection are all local 
operations, nobody outside is able to know how they have actually completed 
the tasks.
\begin{figure}
\epsffile{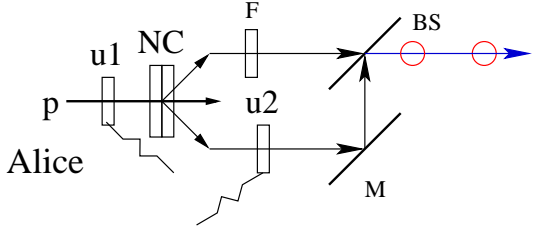}
\vskip 10pt
\caption{ The source of two photons in the same line.
 BS: beam splitter, M: mirror, NC: nonlinear crystals used in SPDC process,
 p: pump light in horizontal polarization, u1: unitary rotator, 
u2: phase shifter. u1 takes the value of 0, $\pi/2$, $\pi/4$ to produce 
emission state
$|11\rangle, |00\rangle, |\phi^+\rangle$, respectively. u2 can be either
$I$ or $\sigma_z$. F interchanges between states $|0\rangle$ and
$|1\rangle$ (i.e., filp between horizontal and vertical polarizations).}\label{source}
\end{figure}  
Intutively, the $collective$ errors could happen more often to qubits
transmitted in the $same$ line. This type of source can be produced by 
figure \ref{source}. Note that the nonlinear crystal there can emmit either
fully entangled state or product state, dependent on
 the polarization of pump beam\cite{spdc}. We can use the following figure to realize the QKD protocol in decoherence free subspace:  
\begin{figure}
\epsffile{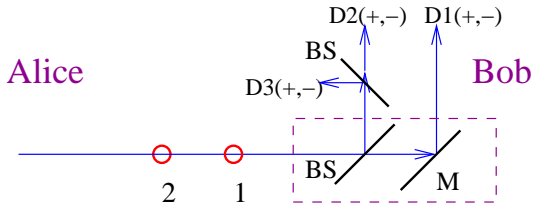}
\vskip 10pt
\caption{ Quantum key distribution in decoherence-free subspace.
The two qubits are now transmitted in the same line. Bob will use  the
results of events of $simtaneous$ clicking of
detectors (D1,D2) or detectors (D1,D3). He discrads all events of 
cliskings at different time or single clicking.   
\label{scheme2} }
\end{figure} 
The security is obvious if we regard the operation by the dashed
rectangular as part of the decoding. This is now a probabilistic decoding
with a rate $75\%$ to discard the code. Consider the case of Shor-Preskill 
protocol where Eve use scheme S' for attacking. Eve detects the right vertical
beam in $|\pm\rangle$ basis and also observes the photon
number of
left vertical beam in the $same$ time. If there is no photon, he blocks
the beam and sends Bob nothing. (Even though there is no photon at that time point, there could be a delayed photon in the same beam. Bob blocks the possible
delayed photon.) If there is one photon, he takes unitary 
transformation U to that photon and sends it to Bob.
 U is defined in step 4 of our protocol. Here Eve has used a very sophisticated
photon detector to detect the photon number but not disturb the polarization
on the left vertical beam.
If Eves takes such operation, to Alice and Bob, they are carrying out 
Shor-Preskill protocol with a
$lossy$ channel with the lossy rate of $75\%$. However, 
we know that Shor-Preskill protocol is secure even the channel is lossy, 
since the corresponding entanglement purification also works with lossy 
channel. Therefore the protocol in fig.\ref{scheme2} must be secure since 
otherwise Shor-Preskill protocol would be insecure with lossy channel.
 
Having considered the QKD protocol with the collective dephasing channel, we 
now consider  the case with collective equatorial rotating channel. 
Such a channel adds collective rotation 
$ \left(
\begin{array}{cc}\cos\theta &\sin\theta\\-\sin\theta & \cos\theta
  \end{array} \right) $ to each individual qubits with $\theta$ value being
random but same to all qubits simultaneously transmitted.
A B92-like protocol was proposed recently\cite{boil}, the protocol
tolerates the collective rotation error\cite{boil}. 
However the security for that protocol is not proven\cite{boil}. 
Also in that protocol, they use 4 qubits
to encode one bit, this will be technically very difficult
and greatly decrease the efficiency because of
the small emission probability of SPDC process. (Even though the probability
can be improved in the future, one still cannot arbitrarily
use large emission probability
because this will cause the un-wanted higher order states.)
 To overcome these drawbacks, we propose a new protocol 
here. Our new protocol has the following advantages: 
1) the security proof is given; 2)we use a 2-qubit state to encode one bit
3) the protocol is BB84-like and the efficiency is further improved compared 
with the one given in ref.\cite{boil}.
Consider the fact that both 2-qubit 
state $|\phi^+\rangle$ and state $|\psi^-\rangle=\frac{1}{\sqrt 2}(|01\rangle
-|10\rangle)$ are unchanged after any collective equatorial rotation
to both qubits. We can use
these two states and their 2 linear superpositions to replace the BB84 state
in QKD. Specificaly, the states transmitted from Alice to Bob are  
randomly chosen from
$\{|\phi^+\rangle, |\psi^-\rangle, \frac{1}{\sqrt 2}(|\phi^++|\psi^-\rangle),
\frac{1}{\sqrt 2}(|\phi^+-|\psi^-\rangle)\}$. Note that
$\frac{1}{\sqrt 2}(|\phi^+\pm|\psi^-\rangle)
=\frac{1}{\sqrt 2}(|0\rangle|\pm\rangle+|1\rangle|\mp\rangle)$. 
Therefore this type of source can be easily made through combining
the type II SPDC emission\cite{para} and  an electrically
driven unitary crystal to make random unitary transformations to one of the two
emitted photons. The state from type II SPDC 
emission is $|\phi^+\rangle$. The random unitary transformations produced 
by the unitary crystal include
unity, joint phase-flip and bit-flip, Hadamard transform and Hadamard transform
with a flipping in $\{|\pm\rangle\}$ basis. Therefore the required
4 2-bit states will be produced randomly with equal distribution.
Unlike protocol in
the case of dephasing channel\cite{walton}, here we don't have to use a
disentangler to generate the source.

According to our theorem, to show the security of this protocol, we need only to show that these 4 2-bit states can be {\it in principle} generated from BB84 states without
any information of the BB84 state
and Bob can recover the original BB84 state on a single qubit by
decoding the 2-bit code. If we can construct the encoding and decoding,
the security of the protocol is equivalent to that of Bb84 protocol.
The encoding indeed exists. To encode, Alice makes the following
unitary transformation
on the BB84 qubit (the first qubit) and the ancilla (the second qubit)
which is in state $|0\rangle$:
\begin{eqnarray}
|00\rangle\longrightarrow \frac{1}{\sqrt 2}(|00\rangle +|11\rangle)
;
|10\rangle \longrightarrow \frac{1}{\sqrt 2}(|01\rangle -|10\rangle)
\label{rota}.
\end{eqnarray}   
In such a way, BB84 state $|\pm\rangle$ with ancilla $|0\rangle$ 
will become $\frac{1}{\sqrt 2}(|\phi^+\rangle\pm |\psi^-\rangle)$.
So in principle, there is indeed an encoding scheme to produce the requested
4 2-bit states from BB84 states. Now we show Bob can recover the original
single qubit state. To do so Bob measuress the first qubit in
$Z$ basis, if he obtains $|0\rangle$, the original state has recovered
on the second qubit, if he obtains $|1\rangle$, he takes
unitary transformation 
$\Sigma=\left(\begin{array}{cc}0 & 1\\-1& 0\end{array}\right)$
to the second qubit and the original state is also recovered up to an
overall phase factor on the second qubit.
However, such a unitary transformation according to the measurement
outcome of the first qubit requires $active$ operations. We can simplify
it to the equivalent $passive$ operation with post-selection. 
Without any loss of generality, a measurement to state $\Sigma |\chi\rangle$ 
in basis 
$\{ |q\rangle\langle q|\}$ is equivalent
to a measurement to $|\chi\rangle$
in basis $\{\Sigma^\dagger |q\rangle\langle q|\Sigma\}$ 
followed by a unitary transformation $\Sigma$. 
This is to say,  Bob takes the $individual$ measurement
to the first qubit
in Z basis and the second 
qubit  in either X basis or Z basis and determine the bit value $b$ according to the 
joint measurement outcome on both qubits:
\begin{eqnarray}
\nonumber \{|0\rangle|0\rangle,|1\rangle|1\rangle,|0\rangle|+\rangle, 
|1\rangle|-\rangle\}\longrightarrow b=0;
\\ \{|0\rangle|1\rangle,|1\rangle|0\rangle,|0\rangle|-\rangle, |1\rangle|+\rangle\}\longrightarrow b=1.
\label{corres2}
\end{eqnarray}
Note that Bob's record on  ``measurement basis'' to each encoded bit
is same as his 
measurement basis to the second qubit. He will discard those results obtained
from a wrong measurement basis through classical communication with Alice.

Moreover, if we exchange the two 
qubits before any measurement done by Bob, all  results of the protocol is
unchanged. That is to say, here Bob can chose any qubit as ``qubit 1''.
Thus in the application Bob just make sure the measurement basis to one qubit
is Z, while the measurement to the other qubit can be either X or Z and
then use the rule of eq.(\ref{corres2}) to determine the bit value. 
The experimantal scheme is given by the following figure \ref{scheme3}.
 Bob will use the results of any 2-fold 
clicking events in $\{(D1,Di),(D3,Di); i=1,2,3,4\}$. He determines the bit 
value according to eq.(\ref{corres2}). Note that  the 2-fold event
(D1,D1) or (D3,D3) means that detector D1 or D3 clicks 2 times at 
{\it different} 
time. Those events of
2-fold clicking which contain neither D1 nor D3 will be discarded. 
\begin{figure}
\epsffile{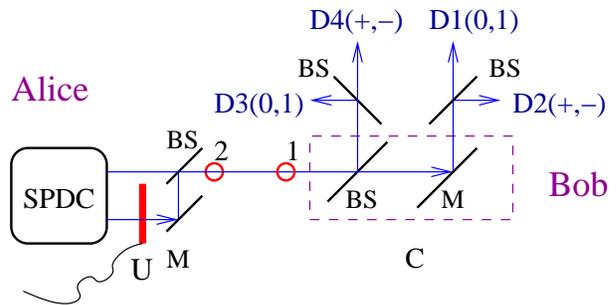}
\vskip 10pt
\caption{ Robust quantum key distribution with collective equatorial rotating
channel. U is a crystal driven electrically to produce random unitary 
transformation selected from $\{I, \Sigma, H, \sigma_zH\}$. $H$ is a Hadamard transform which interchanges $\{|0\rangle, |1\rangle\}$ and $\{|+\rangle,|-\rangle\}$, 
respectively. The devices inside the dashed rectangular are denoted by C.
\label{scheme3} }
\end{figure}  
Foe completness, we give the following protocol with collective rotating channel:
\\{\it Protocol 2.}
{\bf 1:} Alice creates  $(4+\delta)n$
2-qubit quantum codes with each of them randomly chosen from the set $\{|\phi^+\rangle,|\psi^-\rangle,
\frac{1}{\sqrt 2}(|\phi^+\rangle+|\psi^-\rangle), 
\frac{1}{\sqrt 2}(|\phi^+\rangle-|\psi^-\rangle)\}$.
{\bf 2:} For each 2-bit code, if
it is in state $|\phi^+\rangle$ or $|\psi^-\rangle$ she puts down the ``preparation basis'' as 
``Z basis'' otherwise she puts down the ``preparation basis'' as 
``X basis''. 
For those code states of  $|\phi^+\rangle$ or  
$\{\frac{1}{\sqrt 2}(|\phi^+\rangle+|\psi^-\rangle)$, she denotes a bit value 0 ;
for those code states of $|\psi^-\rangle$ or  
$\{\frac{1}{\sqrt 2}(|\phi^+\rangle-|\psi^-\rangle)$, she denotes a bit value 1.
{\bf 3:} Alice sends the 2-qubit codes to Bob.
{\bf 4:} Bob receives the $(4+\delta)n$ 2-qubit codes. To each code,
He measures 
one  qubit (say, qubit 1) in $Z$ basis and  measures the other qubit
(say, qubit 2) in either X basis or Z basis. 
Bob denotes his ``measurement basis'' just the same of his measurement basis
to qubit 2. And he determines the bit value correspondoing to
 each code by eq.(\ref{corres2}).
{\bf 5:} Alice announces her ``preparation basis'' for each codes.
{\bf 6:} Alice and Bob discards those bits on which Bob has
 measured in a basis
       diffrent from Alice's preparation basis.  
         With high probability, there are at
	least $2n$ bits left (if not, abort the protocol).  Alice 
        decides randomly on a set of $n$ bits to use for the protocol, and
        the rest to be check bits. 
{\bf 7:} Alice and Bob announce the values of their check bits.
	If too few of these values agree, they abort the protocol.
{\bf 8:} Alice and Bob distill the final key by using the classical
CSS code\cite{shorpre}.
\\{\bf Acknowledgement:} I thank Prof Imai H for support. 

\end{document}